\documentclass[letter,traditabstract,letterpaper]{aa}  
\usepackage{color}
\usepackage{natbib}
\bibpunct{(}{)}{;}{a}{}{,}
\usepackage{graphicx}
\usepackage{txfonts}

\begin{document}

\titlerunning{A Sample of [C\,{\sc ii}] Clouds Tracing Dense Clouds in Weak FUV Fields}
\authorrunning{Pineda, Velusamy, Langer, Goldsmith,  Li, Yorke}

 \author{ J. L.\,Pineda \and T.\,Velusamy \and W. D. Langer \and
     P. F. Goldsmith \and D. Li. \and H.W. Yorke }

\offprints{J.\,L.\,Pineda \email{Jorge.Pineda@jpl.nasa.gov}}
\institute{Jet Propulsion Laboratory, California Institute of Technology, 4800 Oak Grove Drive, Pasadena, CA 91109-8099, USA}

\title{A Sample of [C\,{\sc ii}] Clouds Tracing Dense Clouds in Weak FUV Fields observed by Herschel\thanks{Herschel is an ESA space observatory with science instruments provided by European-led Principal Investigator consortia and with important participation from NASA.}}

\date{Received / Accepted }

\abstract { 
The [C\,{\sc ii}] fine--structure line at 158\,$\mu$m is an excellent
tracer of the warm diffuse gas in the ISM and the interfaces between
molecular clouds and their surrounding atomic and ionized envelopes.
Here we present the initial results from Galactic Observations of
Terahertz C$^{+}$ ({\bf GOT\,C+}), a Herschel Key Project devoted to
study the [C\,{\sc ii}] fine structure emission in the galactic plane
using the HIFI instrument.  We use the [C\,{\sc ii}] emission together
with observations of CO as a probe to understand the effects of
newly--formed stars on their interstellar environment and characterize
the physical and chemical state of the star-forming gas.  We collected
data along 16 lines--of--sight passing near star forming regions in
the inner Galaxy near longitudes 330\degr and 20\degr.  We
identify fifty-eight [C\,{\sc ii}] components that are associated with
high--column density molecular clouds as traced by $^{13}$CO
emission. We combine [C\,{\sc ii}], $^{12}$CO, and $^{13}$CO
observations to derive the physical conditions of the [C\,{\sc
ii}]--emitting regions in our sample of high--column density clouds
based on comparison with results from a grid of Photon Dominated
Region (PDR) models.  From this unbiased sample, our results suggest
that most of [C\,{\sc ii}] emission originates from clouds with H$_2$
volume densities between $10^{3.5}$ and $10^{5.5}$\,cm$^{-3}$ and weak
FUV strength ($\chi_0=1-10$). We find two regions where our analysis
suggests high densities $>10^{5}$\,cm$^{-3}$ and strong FUV fields
($\chi_0=10^{4}-10^{6}$), likely associated with massive star
formation. We suggest that [C\,{\sc ii}] emission in conjunction with
CO isotopes is a good tool to differentiate between regions of massive
star formation (high densities/strong FUV fields) and regions that are
distant from massive stars (lower densities/weaker FUV fields) along
the line--of--sight. }


\maketitle

\section{Introduction}
\label{sec:introduction}

The study of the processes governing the formation and destruction of
molecular clouds is critical for our understanding of how galaxies have
evolved in our Universe.  In terms of column and local volume
densities only two extreme states of cloud evolution have been
systematically observed: diffuse atomic clouds traced by the 21\,cm
line of H\,{\sc i} \citep[e.g.][]{Kalberla2009} and dense molecular
clouds traced by rotational transitions of CO \citep[e.g.][]{Dame01}.
We know, however, very little about the intermediate phases of cloud
evolution and the interface between diffuse and dense molecular gas.

{\bf G}alactic {\bf O}bservations of {\bf T}erahertz {\bf C$^+$}
({\bf GOT\,C+}), a Herschel Key Project, is devoted to study the
[C\,{\sc ii}] emission in different environments in our Galaxy.  The
survey will observe the [C\,{\sc ii}] 158\,$\mu$m line over a volume
weighted sampling of 500 lines--of--sight (LOS).   Upon completion, it
will provide a database of a few thousand [C\,{\sc ii}]--emitting
clouds distributed over the entire Galactic plane.

The [C\,{\sc ii}] fine structure line at 158\,$\mu$m is an excellent
tracer of the interface between diffuse and dense molecular gas. The
densities and temperatures in this interface allow effective
collisional excitation of this line.  The H\,{\sc i} and ${\rm H}_2$
volume densities are a significant fraction of, or comparable to, the
critical densities for collisional excitation\footnote{Electrons are a
possibly significant collision partner of C$^+$.  However, the
critical electron density for these particles to produce significant
[C\,{\sc ii}] emission is 9.2\,cm$^{-3}$ at $T=100$\,K. 
The density in diffuse regions where the abundance relative to H is
$X(e)\simeq X({\rm C}^+) \simeq 10^{-4}$ is modest
($\sim10^2$\,cm$^{-3}$), while in the denser regions the ionization is
significantly lower.  In either case, the excitation by electrons is
negligible.
} (3.3$\times10^{3}$ and 7.1$\times10^{3}$ cm$^{-3}$
at $T=100$\,K, respectively), the kinetic temperatures are {
$\sim$100\,K}, and the formation of CO is inhibited by limited
shielding against far-ultraviolet (FUV) photons and therefore most of
the gas-phase carbon is in C$^+$ and some C$^0$.

Here we present the first results on the molecular cloud-atomic cloud
interface from the {\bf GOT\,C+} project. During the Herschel Priority
Science and Performance Verification phase, we have collected data
along 5 LOSs near $l=340$\degr\,and 9 LOSs near
$l=20$\degr\,\citep{Velusamy2010}.  The focus of this letter is to
study [C\,{\sc ii}] components towards clouds that have sufficient
column density to have significant $^{13}$CO emission. Such regions
can be considered as dense Photon--Dominated Regions ( or
  photodissociation regions, or PDRs). PDRs are regions where the
chemistry and thermal balance is dominated by the effects of FUV
photons from young stars \citep[][and references
  therein]{HollenbachTielens99}. These data are therefore important
for the study of the stellar feedback of newly formed massive stars in
their progenitor molecular cloud.  We combine the [C\,{\sc ii}] data
with observations of $^{12}$CO and $^{13}$CO from the ATNF Mopra 22-m
telescope to study  58 high--column density PDRs. We use the
[C\,{\sc ii}]/$^{12}$CO and [C\,{\sc ii}]/$^{13}$CO integrated
intensity ratios to constrain physical conditions of the
line--emitting gas comparing with a grid of PDR models.

The Galactic plane has been studied in [C\,{\sc ii}] with low velocity
and spatial resolution observations with COBE \citep{Bennett1994} and
BICE \citep{Nakagawa1998}. The high angular (12\arcsec) and velocity
(0.2\,km\,s$^{-1}$) resolution of the Herschel/HIFI observations allow
us to study for the first time the rich structure of molecular clouds
along the line-of-sight towards the galactic plane.  The Kuiper
Airborne Observatory allowed the study of a handful of H\,{\sc ii}
regions with velocity resolved [C\,{\sc ii}] observations
\citep[e.g.][]{Boreiko1988,Boreiko91}. However, they were limited to
massive star-forming regions with dense and hot PDRs. The sensitivity
of our observations allow us to study for the first time the
population of PDRs in our galaxy that are exposed to weaker FUV
radiation fields.



\section{Observations}
\label{sec:observations}

\begin{figure}[t]
\centering
\includegraphics[width=0.45\textwidth,angle=0]{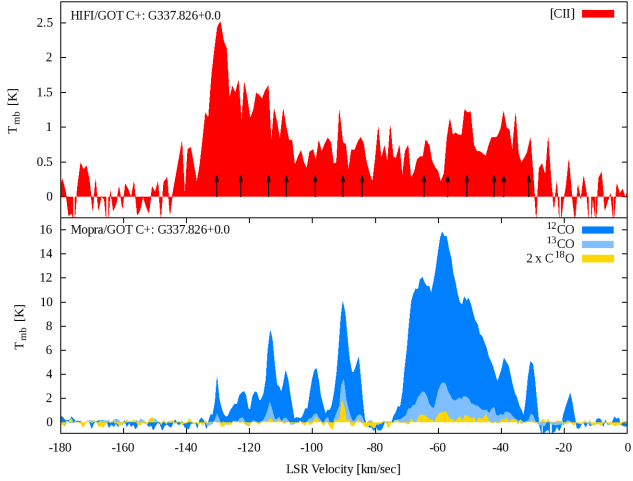}
\caption{[C\,{\sc ii}], $^{12}$CO,$^{13}$CO, and C$^{18}$O spectra
towards the line--of--sight G337.826+0.0. The arrows indicate the
[C\,{\sc ii}] components that have $^{13}$CO
counterparts. }\label{fig:example_los}
\end{figure}

\subsection{Herschel Observations}\label{sec:c-sc-ii}

We observed the [C\,{\sc ii}] $^2$P$_{3/2} \to ^2$P$_{1/2}$ line at
1900.5469\,GHz towards 16 LOSs in the Galactic plane with the HIFI
\citep{deGraauw2010} instrument on board  the Herschel space
observatory \citep{Pilbratt2010}. We refer to \citet{Velusamy2010} for
more details about the [C\,{\sc ii}] observations. In
Figure~\ref{fig:example_los} we show sample LOS spectra of [C\,{\sc
    ii}], $^{12}$CO, $^{13}$CO, and C$^{18}$O.

\subsection{Mopra Observations}\label{sec:mopra-observations}

\begin{figure*}[t]
\centering
\includegraphics[width=0.95\textwidth,angle=0]{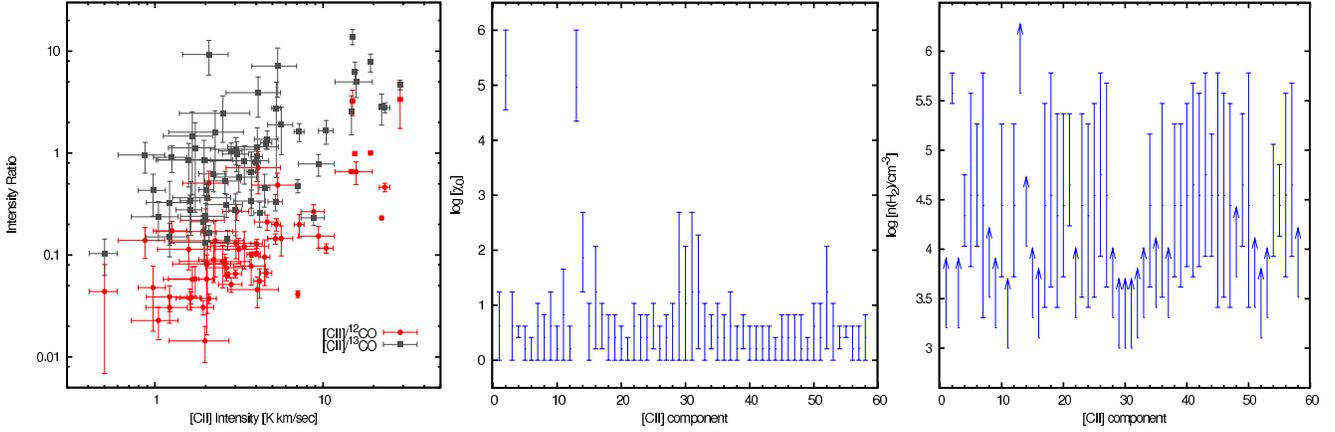}
\caption{({\it left panel}) The observed [C\,{\sc ii}]/$^{12}$CO and
[C\,{\sc ii}]/$^{13}$CO integrated intensity ratios for our sample of
[C\,{\sc ii}] components associated with high--column density
molecular clouds.  The ratios are calculated from integrated
intensities in units of K\,km\,s$^{-1}$.  The horizontal error bars
are the uncertainties in the determination of the [C\,{\sc ii}]
integrated intensities. The vertical error bars are the uncertainties
in the line ratios which are derived from error propagation.  ({\it
central and right panels}) Results of the comparison between [C\,{\sc
ii}]/$^{12}$CO and [C\,{\sc ii}]/$^{13}$CO ratios for all identified
[C\,{\sc ii}] components and the PDR model grid showing the
constrained ranges in FUV radiation field ({\it central panel}) and
H$_2$ volume density ({\it right panel}). The arrows indicate lower
limits to the H$_2$ volume density. }
\label{fig:results_pdr}
\end{figure*}

We observed the $J = 1 \to 0$ transitions of $^{12}$CO, $^{13}$CO, and
C$^{18}$O toward the observed [C\,{\sc ii}] LOSs.  These observations
are part of a survey of all {\bf GOT\,C+} positions towards the inner
Galaxy between $l=-175.5$\degr and $l=56.8$\degr conducted at the ATNF
Mopra Telescope. The angular resolution of these observations is
33\arcsec.  Typical system temperatures were 600, 300, and 250\,K for
$^{12}$CO, $^{13}$CO, and C$^{18}$O, respectively.  To convert from
antenna to main--beam temperature scale we use a main-beam efficiency
of 0.42 \citep{Ladd05}. All lines were observed simultaneously using
the MOPS spectrometer in zoom mode. The spectra were smoothed in
velocity to 0.8\,km s$^{-1}$ for $^{12}$CO and $^{13}$CO and to
1.6\,km\,s$^{-1}$ for C$^{18}$O. Typical rms noise is 0.6\,K for
$^{12}$CO and 0.1\,K for both $^{13}$CO and C$^{18}$O.  We checked
pointing accuracy every 60 minutes using the closest and brightest SiO
maser.

\section{[C\,{\sc ii}] components associated with molecular clouds}
\label{sec:cii-comp-assoc}

We identify a total of 146 [C\,{\sc ii}] velocity components in the
observed LOSs.  From this data set we identify components
that are associated with high--column density molecular gas by looking for $^{13}$CO
counterparts. We identified most of the high--$^{13}$CO column density
[C\,{\sc ii}] components by fitting Gaussian functions defined by
fitting the corresponding $^{13}$CO spectra. The only exception was
G337.826+0.0 for which we calculated the integrated intensity by
determining the area within the FHWM of the $^{13}$CO emission, as
this line--of--sight shows complex velocity structure.  Based on the
$^{13}$CO line parameters we identify 58 [C\,{\sc ii}] components
associated with dense molecular gas.  All of them also show $^{12}$CO
emission while 12 show C$^{18}$O emission.  The remaining diffuse
atomic and/or diffuse molecular [C\,{\sc ii}]--emitting clouds that do
not have $^{13}$CO counterparts are discussed by \citet{Langer2010}
and \citet{Velusamy2010}.

In the left panel of Figure~\ref{fig:results_pdr}, we summarize the
observed characteristics by plotting the [C\,{\sc ii}]/$^{12}$CO and
[C\,{\sc ii}]/$^{13}$CO integrated intensity ratios for the identified
components as a function of the [C\,{\sc ii}] integrated intensity.
The ratios are calculated from integrated intensities in units of
K\,km\,s$^{-1}$. The mean value and standard deviation are 0.29 and
0.6 for the [C\,{\sc ii}]/$^{12}$CO integrated intensity ratio and
1.75 and 2.54 for [C\,{\sc ii}]/$^{13}$CO.  The ratios vary over 2
orders of magnitude suggesting a wide range of physical conditions in
our sample.


We use the [C\,{\sc ii}]/$^{12}$CO and [C\,{\sc ii}]/$^{13}$CO
integrated intensity ratios to constrain the physical conditions of
the line--emitting gas.  The $^{12}$CO emission, which becomes
optically thick quickly after a modest fraction of the gas--phase
carbon is converted to CO, is not very sensitive to the FUV radiation
field, as the temperature at the C$^+$/C$^0$/CO transition layer is also
insensitive to this quantity \citep{Wolfire1989,
Kaufman99}. Therefore, the [C\,{\sc ii}]/$^{12}$CO ratio is determined
by the column density of C$^+$ and the temperature at the surface of
the PDR, which are in turn dependent on the FUV radiation field and
H$_2$ density.  The [C\,{\sc ii}]/$^{13}$CO ratio is proportional to
the ratio between the C$^+$ and $^{13}$CO column densities. It
therefore gives, provided that extra constraints on the total column
of material are available and that there are no significant variations
of the FUV field within the beam, a constraint on the location of the
C$^+$/C$^0$/CO transition layer which in turn depends on the strength of
the FUV field and H$_2$ density.

\section{Comparison with  PDR model calculations}
\label{sec:comparison-with-pdr}

We compare the observed [C\,{\sc ii}]/$^{12}$CO and [C\,{\sc
ii}]/$^{13}$CO integrated intensity ratios with the results of a PDR
model grid in order to constrain physical conditions of the [C\,{\sc
ii}]--emitting clouds. 

The model grid was calculated using the KOSMA--$\tau$ PDR model
\citep{Stoerzer96,Roellig06} which is available online\footnote{{\tt
http://hera.ph1.uni-koeln.de/$\sim$pdr/ }}.  The model provides a
self--consistent solution of the chemistry and thermal balance of a
spherical cloud, with a truncated density profile, which is
illuminated isotropically by a FUV radiation field.  The density
distribution has the form, $n(r)$=$n_{s}(r/r_c)^{-1.5}$ for 0.2$r_c$
$\leq$ $ r \leq r_c$ and a constant density of
$n(r)=n_s$(0.2)$^{-1.5}$ in the central region of the cloud ($r <
0.2r_c$). Here $r_c$ is the cloud radius and $n_{s}$ is the density at
the cloud surface. Note that with a power--law index of 1.5, the
average density of the clump is about twice the density at the cloud
surface. The line intensities are calculated using a non--LTE
radiative transfer code by \cite{Gierens92}. Each model is
characterized by the clump mass, the density at the cloud surface, and
strength of the FUV field. The clump mass ranges from $10^{-2}$ to
$100$\,M$_\odot$, the density at the cloud surface from 10$^3$ to
10$^6$\,cm$^{-3}$, and the strength of the FUV field from $\chi_{\rm
0}=1$ to 10$^6$ (in units of the \citealt{Draine78} field\footnote{
The average FUV intensity of the local ISM is
2.2$\times10^{-4}$\,erg\,cm$^{-2}$\,s$^{-1}$\,sr$^{-1}$
\citep{Draine78}. Note that the Draine field is isotropic (i.e. a
given point is illuminated from 4$\pi$ steradians) while the surface
of the clouds considered here are only illuminated from 2$\pi$
steradians. Therefore the rate of photoreactions at the cloud surface
are half of what they would be with the Draine field in free space.
}).  We do not use H\,{\sc i} and C$^{18}$O observations to constrain
our solutions as model grids involving their emission are not
available.

By using a spherically symmetric model we assume that the cloud
spatial structure can be described by an ensemble of clumps with sizes
much smaller than the resolution of our observations. Additionally, we
assume that each clump in this ensemble has the same mass and density,
and that the [C\,{\sc ii}]/$^{12}$CO and [C\,{\sc ii}]/$^{13}$CO line
ratios can be estimated using the line ratios of a single clump of
that mass and density.  Therefore, the comparison with the PDR model
grid provides the typical  incident FUV field, mass, and density
of the regions that dominate the observed line ratios.  An even more
realistic model considers clumps following the distribution of masses
and sizes observed in many molecular clouds
\citep[e.g.][]{Zielinsky2000,Cubick2008}.

\begin{figure}[t]
\centering
\includegraphics[width=0.4\textwidth,angle=0]{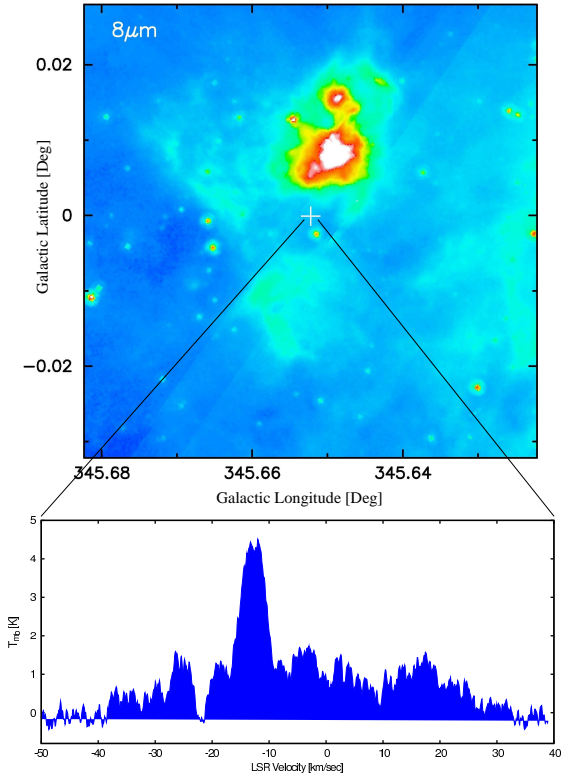}
\caption{Example of  [C\,{\sc ii}] emission  associated with a
massive star forming region. The line--of--sight G345.65+0.0 passes
near several bright H\,{\sc ii} regions as shown in the Spitzer 8\,$\mu$m
image. In this line--of--sight we find a bright [C\,{\sc ii}]
velocity component at -12.9 km\,s$^{-1}$ which our analysis suggests arises
from a region with high density ($\sim10^{5.7}$\,cm$^{-3}$) and strong
FUV field ($\chi_0=10^{4.5}-10^6$).  }\label{fig:spectra2}
\end{figure}

The central and right panels in Figure~\ref{fig:results_pdr} show a
summary of the constrained H$_2$ volume densities and FUV radiation
fields for our sample.  We consider models with chi-squared $\chi^2$
smaller than 1.1$\chi^2_{\rm min}$.  We find two [C\,{\sc ii}]
components with high volume densities ($>10^5$\,cm$^{-3}$) and strong
FUV fields (between $ \chi_{\rm 0} = 10^4-10^6$).  Both regions
are characterized by [C\,{\sc ii}]/CO integrated intensity ratios that
are larger than 1 (c.f. Orion has a [C\,{\sc ii}]/$^{12}$CO ratio of
1.36; \citealt{Crawford1985}).  
We show an image and [C\,{\sc ii}]  spectrum of one such source in 
Figure~\ref{fig:spectra2}.  The remaining components have lower volume
densities between $10^{3.5}-10^{5.5}$\,cm$^{-3}$.   Six of those
could have a strength of FUV field as high as 100 while the majority
(51 components) have FUV fields between 1 and 10.   For all
components the comparison with the PDR model grid suggests clump
masses that are larger than 1\,M$_\odot$.   Note that due to the
limited spatial coverage of the observations presented here, the
distribution of physical conditions is not smooth.  We will obtain a
much better sampling of the distribution of physical conditions in
velocity components distributed over the entire galactic plane with
the completed {\bf GOT C+} survey.

The large number of components with low--FUV field is a result of the
low observed [C\,{\sc ii}]/$^{12}$CO ratios of about 0.1. Such values
of the [C\,{\sc ii}]/$^{12}$CO ratio are expected for $ \chi_0 <
10^{3}$ over a large range of H$_2$ volume densities (see
e.g. Figure\,9 in \citealt{Kaufman99}). The [C\,{\sc ii}]/$^{13}$CO
ratio provides an additional constraint on the FUV field.  The
majority of the observed components have small ratios that suggest a
large column density of $^{13}$CO relative to that of C$^+$. This
result is suggestive of a C$^+$/C$^0$/CO transition occurring close to the
surface of the cloud which is a result of either high densities or
weak FUV fields.  Note, however, that using [C\,{\sc ii}] and
$^{13}$CO to constrain the location of the C$^+$/C$^0$/CO transition
requires extra constraints on the total column density of material
thoughout the clump which in turns depends on the assumed clump
surface density and mass. These two quantities are not well
constrained in the analysis presented here. Additionally, it requires
that there are no significant variations of the FUV field within the
beam as shielded clumps might contribute significantly to the
$^{13}$CO emission but little to that of [C\,{\sc ii}].  The H$_2$
volume density for individual velocity components can be better
determined using observations of the 609$\mu$m and 370$\mu$m
transitions of neutral carbon, which have been used to constrain the
temperature and density at the C$^+$/C$^0$/CO transition region in PDRs
\citep[e.g.][]{Stutzki1997}. The [C\,{\sc ii}] to bolometric infrared
flux is also useful to constrain the FUV field
\citep{Wolfire1989,Kaufman99} but is only useful towards LOSs with a
single velocity component.

\section{Discussion}
\label{sec:discussion}


We have found that most of the [C\,{\sc ii}] velocity components
considered here are associated with regions that are exposed to weak
FUV radiation field ($\chi_0 <10$) and therefore are away from OB
associations. PDRs exposed to weak FUV radiation fields have been
studied in a few sources using [C\,{\sc i}] emission
\citep[e.g.][]{Maezawa1999,Bensch06,Pineda07a} but never observed with
velocity--resolved [C\,{\sc ii}] before Herschel/HIFI.


\citet{Cubick2008} suggested that most of the [C\,{\sc ii}] in our
Galaxy originates from a clumpy medium exposed to a FUV field of
$\chi_0=10^{1.8}$, which is larger than the upper limit determined for
the majority of the observed components.  Note, however, that their
model does not consider emission arising from diffuse clouds. From our
observed LOSs, we find that about 56\% of the total detected [C\,{\sc
ii}] emission arises from low--column density regions (without significant
$^{13}$CO emission; \citealt{Langer2010,Velusamy2010}) while 44\% is
emitted from dense PDRs.  Nevertheless, the moderate FUV field
predicted by \citet{Cubick2008} might suggest that the predominance of
low--FUV radiation field regions observed in our limited sample
(covering less than 2\% of the entire {\bf GOT C+} survey) might hold
for the entire Galaxy.

We found two regions where our analysis suggests high densities
($>10^5$\,cm$^{-3}$) and strong FUV fields (between $\chi_{\rm 0} =
10^4$ and $10^6$).  These regions are likely associated with massive
star formation.  This  conclusion is a
result of the elevated [C\,{\sc ii}]/CO ratio observed towards these
regions.  This identification suggests that the [C\,{\sc ii}]/CO
ratio is good tracer of the location of massive star formation regions
in the galaxy.  [C\,{\sc ii}] observations will therefore provide an
alternative method to determine the distribution of massive star
forming regions in the galaxy \citep[e.g.][]{Bronfman2000}.   Note
that velocity resolved observations are crucial for the interpretation
of the [C\,{\sc ii}]/CO ratio. In our observed LOSs we have found
velocity components showing [C\,{\sc ii}] emission but no CO as well
as components showing CO but no [C\,{\sc ii}]. Velocity unresolved
observations would have given a distorted value of the [C\,{\sc
ii}]/CO ratio that would result in an incorrect interpretation of the
physical conditions of the line--emitting gas.


%

\section{Conclusions}
\label{sec:conclusions}

We have presented velocity--resolved observations of [C\,{\sc ii}]
towards  16 LOSs located near $l=340$\degr\,and $l=20$\degr\,in
the Galactic plane using the HIFI instrument on board  the Herschel
space observatory. We identified a total of  146 different
[C\,{\sc ii}] velocity components.  In this letter we analyzed a
sample of 58 components that are associated with high--column density
molecular gas as traced by $^{13}$CO emission. These components
contribute 44\% of the total observed [C\,{\sc ii}] emission 
implying a significantly larger amount of [C\,{\sc ii}] emission
originating in the diffuse ISM than from star forming environments.
We have compared the [C\,{\sc ii}]/$^{12}$CO and [C\,{\sc
ii}]/$^{13}$CO integrated intensity ratios with a PDR model grid to
constrain the strength of the FUV field and the H$_2$ volume density
in this sample. We find two clouds for which our analysis suggests
high densities ($>10^{5}$ cm$^{-3}$) and strong FUV fields ($\chi_{\rm
0} = 10^{4}-10^{6}$), likely associated with massive star
formation. The majority of the observed components, however, have
modest densities ($10^{3.5}-10^{5.5}$\,cm$^{-3}$) and weaker FUV
fields ($\chi_0=1-10$). Although the population of clouds with these
conditions is likely where most of the [C\,{\sc ii}] emission
originates in our Galaxy, their properties are largely unexplored.
The {\bf GOT\,C+} survey will provide a few thousand clouds
distributed in the Galactic plane and therefore will allow us to
characterize this population of intermediate clouds.

\begin{acknowledgements}

We would like to thank the referee David Hollenbach for his comments
and suggestions that significantly improved this letter. This work was
performed by the Jet Propulsion Laboratory, California Institute of
Technology, under contract with the National Aeronautics and Space
Administration.  We thank the staffs of the ESA and NASA Herschel
Science Centers for their help.  The Mopra Telescope is managed by the
Australia Telescope, which is funded by the Commonwealth of Australia
for operation as a National Facility by the CSIRO.
\end{acknowledgements}

\bibliography{/home/jpineda/latex/papers}

\begin{thebibliography}{26}
\expandafter\ifx\csname natexlab\endcsname\relax\def\natexlab#1{#1}\fi

\bibitem[{{Bennett} {et~al.}(1994){Bennett}, {Fixsen}, {Hinshaw}, {Mather},
  {Moseley}, {Wright}, {Eplee}, {Gales}, {Hewagama}, {Isaacman}, {Shafer}, \&
  {Turpie}}]{Bennett1994}
{Bennett}, C.~L., {Fixsen}, D.~J., {Hinshaw}, G., {et~al.} 1994, \apj, 434, 587

\bibitem[{{Bensch}(2006)}]{Bensch06}
{Bensch}, F. 2006, \aap, 448, 1043

\bibitem[{{Boreiko} \& {Betz}(1991)}]{Boreiko91}
{Boreiko}, R.~T. \& {Betz}, A.~L. 1991, \apjl, 380, L27

\bibitem[{{Boreiko} {et~al.}(1988){Boreiko}, {Betz}, \&
  {Zmuidzinas}}]{Boreiko1988}
{Boreiko}, R.~T., {Betz}, A.~L., \& {Zmuidzinas}, J. 1988, \apjl, 325, L47

\bibitem[{{Bronfman} {et~al.}(2000){Bronfman}, {Casassus}, {May}, \&
  {Nyman}}]{Bronfman2000}
{Bronfman}, L., {Casassus}, S., {May}, J., \& {Nyman}, L. 2000, \aap, 358, 521

\bibitem[{{Crawford} {et~al.}(1985){Crawford}, {Genzel}, {Townes}, \&
  {Watson}}]{Crawford1985}
{Crawford}, M.~K., {Genzel}, R., {Townes}, C.~H., \& {Watson}, D.~M. 1985,
  \apj, 291, 755

\bibitem[{{Cubick} {et~al.}(2008){Cubick}, {Stutzki}, {Ossenkopf}, {Kramer}, \&
  {R{\"o}llig}}]{Cubick2008}
{Cubick}, M., {Stutzki}, J., {Ossenkopf}, V., {Kramer}, C., \& {R{\"o}llig}, M.
  2008, \aap, 488, 623

\bibitem[{{Dame} {et~al.}(2001){Dame}, {Hartmann}, \& {Thaddeus}}]{Dame01}
{Dame}, T.~M., {Hartmann}, D., \& {Thaddeus}, P. 2001, \apj, 547, 792

\bibitem[{{de Graauw} {et~al.}(2010){de Graauw}, {Helmich}, {Phillips},
  {Stutzki}, {Caux}, {Whyborn}, {Dieleman}, {Roelfsema}, {Aarts}, {Assendorp},
  {Bachiller}, {Baechtold}, {Barcia}, {Beintema}, {Belitsky}, {Benz}, {Bieber},
  {Boogert}, {Borys}, {Bumble}, {Ca{\"i}s}, {Caris}, {Cerulli-Irelli},
  {Chattopadhyay}, {Cherednichenko}, {Ciechanowicz}, {Coeur-Joly}, {Comito},
  {Cros}, {de Jonge}, {de Lange}, {Delforges}, {Delorme}, {den Boggende},
  {Desbat}, {Diez-Gonz{\'a}lez}, {di Giorgio}, {Dubbeldam}, {Edwards},
  {Eggens}, {Erickson}, {Evers}, {Fich}, {Finn}, {Franke}, {Gaier}, {Gal},
  {Gao}, {Gallego}, {Gauffre}, {Gill}, {Glenz}, {Golstein}, {Goulooze},
  {Gunsing}, {G{\"u}sten}, {Hartogh}, {Hatch}, {Higgins}, {Honingh}, {Huisman},
  {Jackson}, {Jacobs}, {Jacobs}, {Jarchow}, {Javadi}, {Jellema}, {Justen},
  {Karpov}, {Kasemann}, {Kawamura}, {Keizer}, {Kester}, {Klapwijk}, {Klein},
  {Kollberg}, {Kooi}, {Kooiman}, {Kopf}, {Krause}, {Krieg}, {Kramer},
  {Kruizenga}, {Kuhn}, {Laauwen}, {Lai}, {Larsson}, {Leduc}, {Leinz}, {Lin},
  {Liseau}, {Liu}, {Loose}, {L{\'o}pez-Fernandez}, {Lord}, {Luinge}, {Marston},
  {Mart{\'{\i}}n-Pintado}, {Maestrini}, {Maiwald}, {McCoey}, {Mehdi}, {Megej},
  {Melchior}, {Meinsma}, {Merkel}, {Michalska}, {Monstein}, {Moratschke},
  {Morris}, {Muller}, {Murphy}, {Naber}, {Natale}, {Nowosielski}, {Nuzzolo},
  {Olberg}, {Olbrich}, {Orfei}, {Orleanski}, {Ossenkopf}, {Peacock}, {Pearson},
  {Peron}, {Phillip-May}, {Piazzo}, {Planesas}, {Rataj}, {Ravera}, {Risacher},
  {Salez}, {Samoska}, {Saraceno}, {Schieder}, {Schlecht}, {Schl{\"o}der},
  {Schm{\"u}lling}, {Schultz}, {Schuster}, {Siebertz}, {Smit}, {Szczerba},
  {Shipman}, {Steinmetz}, {Stern}, {Stokroos}, {Teipen}, {Teyssier}, {Tils},
  {Trappe}, {van Baaren}, {van Leeuwen}, {van de Stadt}, {Visser}, {Wildeman},
  {Wafelbakker}, {Ward}, {Wesselius}, {Wild}, {Wulff}, {Wunsch}, {Tielens},
  {Zaal}, {Zirath}, {Zmuidzinas}, \& {Zwart}}]{deGraauw2010}
{de Graauw}, T., {Helmich}, F.~P., {Phillips}, T.~G., {et~al.} 2010, \aap, 518,
  L6+

\bibitem[{{Draine}(1978)}]{Draine78}
{Draine}, B.~T. 1978, \apjs, 36, 595

\bibitem[{{Gierens} {et~al.}(1992){Gierens}, {Stutzki}, \&
  {Winnewisser}}]{Gierens92}
{Gierens}, K.~M., {Stutzki}, J., \& {Winnewisser}, G. 1992, \aap, 259, 271

\bibitem[{{Hollenbach} \& {Tielens}(1999)}]{HollenbachTielens99}
{Hollenbach}, D.~J. \& {Tielens}, A.~G.~G.~M. 1999, Reviews of Modern Physics,
  71, 173

\bibitem[{{Kalberla} \& {Kerp}(2009)}]{Kalberla2009}
{Kalberla}, P.~M.~W. \& {Kerp}, J. 2009, \araa, 47, 27

\bibitem[{{Kaufman} {et~al.}(1999){Kaufman}, {Wolfire}, {Hollenbach}, \&
  {Luhman}}]{Kaufman99}
{Kaufman}, M.~J., {Wolfire}, M.~G., {Hollenbach}, D.~J., \& {Luhman}, M.~L.
  1999, \apj, 527, 795

\bibitem[{{Ladd} {et~al.}(2005){Ladd}, {Purcell}, {Wong}, \&
  {Robertson}}]{Ladd05}
{Ladd}, N., {Purcell}, C., {Wong}, T., \& {Robertson}, S. 2005, PASA, 22, 62

\bibitem[{{Langer} {et~al.}(2010){Langer}, {Velusamy}, {Pineda}, {Goldsmith},
  {Li}, \& {Yorke}}]{Langer2010}
{Langer}, W.~D., {Velusamy}, T., {Pineda}, J.~L., {et~al.} 2010,
  ArXiv:1007.3048

\bibitem[{{Maezawa} {et~al.}(1999){Maezawa}, {Ikeda}, {Ito}, {Saito},
  {Sekimoto}, {Yamamoto}, {Tatematsu}, {Arikawa}, {Aso}, {Noguchi}, {Shi},
  {Miyazawa}, {Saito}, {Ozeki}, {Fujiwara}, {Ohishi}, \&
  {Inatani}}]{Maezawa1999}
{Maezawa}, H., {Ikeda}, M., {Ito}, T., {et~al.} 1999, \apjl, 524, L129

\bibitem[{{Nakagawa} {et~al.}(1998){Nakagawa}, {Yui}, {Doi}, {Okuda}, {Shibai},
  {Mochizuki}, {Nishimura}, \& {Low}}]{Nakagawa1998}
{Nakagawa}, T., {Yui}, Y.~Y., {Doi}, Y., {et~al.} 1998, \apjs, 115, 259

\bibitem[{{Pilbratt} {et~al.}(2010){Pilbratt}, {Riedinger}, {Passvogel},
  {Crone}, {Doyle}, {Gageur}, {Heras}, {Jewell}, {Metcalfe}, {Ott}, \&
  {Schmidt}}]{Pilbratt2010}
{Pilbratt}, G.~L., {Riedinger}, J.~R., {Passvogel}, T., {et~al.} 2010, \aap,
  518, L1+

\bibitem[{{Pineda} \& {Bensch}(2007)}]{Pineda07a}
{Pineda}, J.~L. \& {Bensch}, F. 2007, \aap, 470, 615

\bibitem[{{R{\"o}llig} {et~al.}(2006){R{\"o}llig}, {Ossenkopf}, {Jeyakumar},
  {Stutzki}, \& {Sternberg}}]{Roellig06}
{R{\"o}llig}, M., {Ossenkopf}, V., {Jeyakumar}, S., {Stutzki}, J., \&
  {Sternberg}, A. 2006, \aap, 451, 917

\bibitem[{{St\"orzer} {et~al.}(1996){St\"orzer}, {Stutzki}, \&
  {Sternberg}}]{Stoerzer96}
{St\"orzer}, H., {Stutzki}, J., \& {Sternberg}, A. 1996, \aap, 310, 592

\bibitem[{{Stutzki} {et~al.}(1997){Stutzki}, {Graf}, {Haas}, {Honingh},
  {Hottgenroth}, {Jacobs}, {Schieder}, {Simon}, {Staguhn}, {Winnewisser},
  {Martin}, {Peters}, \& {McMullin}}]{Stutzki1997}
{Stutzki}, J., {Graf}, U.~U., {Haas}, S., {et~al.} 1997, \apjl, 477, L33+

\bibitem[{{Velusamy} {et~al.}(2010){Velusamy}, {Langer}, {Pineda}, {Goldsmith},
  {Li.}, \& {Yorke}}]{Velusamy2010}
{Velusamy}, T., {Langer}, W.~D., {Pineda}, J.~L., {et~al.} 2010,
  ArXiv:1007.3338

\bibitem[{{Wolfire} {et~al.}(1989){Wolfire}, {Hollenbach}, \&
  {Tielens}}]{Wolfire1989}
{Wolfire}, M.~G., {Hollenbach}, D., \& {Tielens}, A.~G.~G.~M. 1989, \apj, 344,
  770

\bibitem[{{Zielinsky} {et~al.}(2000){Zielinsky}, {Stutzki}, \&
  {St{\"o}rzer}}]{Zielinsky2000}
{Zielinsky}, M., {Stutzki}, J., \& {St{\"o}rzer}, H. 2000, \aap, 358, 723

\end{thebibliography}
\bibliographystyle{aa}

%
\clearpage
\end{document}